\documentclass[11pt,onecolumn]{article}

\usepackage{acronym} 
\usepackage{amsmath}
\usepackage{amssymb}
\usepackage{graphicx}
\usepackage{caption}
\usepackage{natbib}
\usepackage{subcaption}
\usepackage{color}
\usepackage{changepage}
\usepackage{algorithm}
\usepackage{algpseudocode}
\usepackage{hyperref}
\usepackage{fullpage}

\acrodef{ngs}[NGS]{next generation sequencing}
\acrodef{mmd}[MMD]{maximum mean discrepancy}
\acrodef{mch}[mCH]{non-CpG methylation}
\acrodef{rrbs}[RRBS]{reduced representation bisulfite sequencing}
\acrodef{wgbs}[WGBS]{whole genome bisulfite sequencing}
\acrodef{cgis}[CGIs]{CpG regions}
\acrodef{dmr}[DMR]{differentially methylated region}
\acrodefplural{dmr}[DMRs]{differentially methylated regions}
\acrodef{dmc}[DMC]{differentially methylated cytosine}
\acrodefplural{dmc}[DMCs]{differentially methylated cytosines}
\acrodef{m3d}[M$^3$D]{Maximum Mean Methylation Discrepancy}
\acrodef{rkhs}[RKHS]{reproducing kernel Hilbert space}
\acrodef{iid}[i.i.d.]{independently and identically distributed}
\acrodef{rbf}[RBF]{radial basis function}
\acrodef{fdr}[FDR]{false discovery rate}
\acrodefplural{fdr}[FDRs]{false discovery rates}
\acrodef{esc}[ESC]{embryonic stem cell}
\acrodefplural{esc}[ESCs]{embryonic stem cells}
\acrodef{tss}[TSS]{transcription start site}
\acrodef{fpkm}[FPKM]{fragments per kilobase of transcript per million mapped reads}
\acrodef{go}[GO]{gene ontology}
\acrodef{fet}[FET]{Fisher's Exact Test}

\renewcommand*{\thesection}{\arabic{section}}

\title{\vspace{-15mm}\fontsize{24pt}{10pt}\selectfont\textbf{M$^3$D: a kernel-based test for shape changes in methylation profiles}} 

\author{
\large
\textsc{Tom Mayo\,$^{1}$, Gabriele Schweikert\,$^{1,2}$ and Guido Sanguinetti\,${^1}$}\\[2mm] 
\normalsize $^{1}$IANC, School of Informatics, University of Edinburgh\\
$^{2}$Wellcome Trust Centre for Cell Biology, University of Edinburgh\\ 
\vspace{-5mm}
}
\date{}

\begin{document}

   \maketitle

\begin{abstract}

DNA methylation is an intensely studied epigenetic mark implicated in many biological processes of direct clinical relevance. While sequencing based technologies are increasingly allowing high resolution measurements of DNA methylation, statistical modelling of such data is still challenging. In particular, statistical identification of \acp{dmr} across different conditions poses unresolved challenges in accounting for spatial correlations within the statistical testing procedure.

We propose a non-parametric, kernel-based method, M$^3$D, to detect higher-order changes in methylation profiles, such as shape, across pre-defined regions. The test statistic explicitly accounts for differences in coverage levels between samples, thus handling in a principled way a major confounder in the analysis of methylation data. Empirical tests on real and simulated data sets show an increased power compared to established methods, as well as considerable robustness with respect to coverage and replication levels.

Code is implemented in R, and is freely available via Bioconductor package M3D.

\end{abstract}

\acresetall

\section{Introduction}

DNA methylation is an epigenetic mark associated with many fundamental biological processes of direct clinical relevance, such as imprinting, retrotransposon silencing and cell differentiation (\citealp{gopalakrishnan2008dna,laurent2010dynamic}). Methylation occurs when a methyl group is attached to a cytosine. In mammals, methylation is observed predominantly in the CpG context and, consequently, studies tend to focus on these loci. The canonical understanding is that methylation of \ac{cgis} in promoter regions is associated with gene silencing, however, recent studies have shown that CpG methylation correlates with gene expression in a more complex and context-dependent manner (\citealp{varley2013dynamic}). Methylation profiles are altered in many diseases, most notably cancer (\citealp{das2004dna,sharma2010epigenetics}), and as such epigenetic therapies are being developed which specifically target methylation (\citealp{yang2010targeting}). 

Bisulfite treatment of DNA followed by next generation sequencing provides quantitative methylation data with base pair resolution. Unmethylated cytosines are deaminated into uracils, which amplify as thymines during PCR (\citealp{krueger2012dna}). Reads are then aligned to a reference genome, permitting changes of C to T. The resulting counts of cytosine and thymine at registered cytosine loci form the basis of further analysis. This general procedure has been adapted in various ways, with \ac{rrbs} being one of the most widely used. \ac{rrbs} involves using a restriction enzyme such as MspI (or TaqI) to cleave the DNA at CCGG (or TCGA) loci and selecting short reads for sequencing (\citealp{gu2011preparation}). This results in greater coverage of CpG dense regions at lower cost.

Several methods have been proposed to statistically test for  \acp{dmr}. {Almost all of these methods perform a search for \acp{dmr} by testing individual cytosines followed by a post-hoc aggregation procedure. Early methylation studies used \ac{fet} to identify \acp{dmc} (\citealp{li2010dna,challen2012dnmt3a}). BSmooth (\citealp{hansen2012bsmooth}), one of the most widely used methods, performs local likelihood smoothing to generate methylation profiles for each sample, before testing individual locations in the profiles to identify \acp{dmc}. More recent methods, such as BiSeq (\citealp{hebestreit2013detection}) and methylSig (\citealp{park2014methylsig}), also employ local smoothing, together with a beta-binomial model of methylation at individual cytosines; both of these methods then aggregate the results of tests at individual loci to compute a measure of significance for \acp{dmr}. The beta-binomial {method} models biological variability at each cytosine location and hence requires a high replication level to achieve power. Coverage can also be problematic, since low coverage precludes statistical significance and high coverage can lead to over-confidence in calling \acp{dmr}, although the latter effect can be ameliorated by having a larger number of replicates. For instance, methylSig requires a minimum of 3 replicates per group and ignores loci which are covered by fewer than 10 reads by default. 
A recent method, MAGI \citep{baumann2014magi}, takes a different approach by testing directly for \acp{dmr}, rather than computing region-wide measures of significance from tests of individual cytosines. MAGI assumes the availability of genome-wide decision boundary methylation levels (which can be determined either from annotation or in a data driven fashion). Methylation levels at each cytosine are then given a binary representation based on whether they exceed the decision boundary, and a single \ac{fet} is performed over each region by counting how many cytosines have changed state.}

\begin{figure}[!tpb]
\centerline{\includegraphics[width=0.5\textwidth]{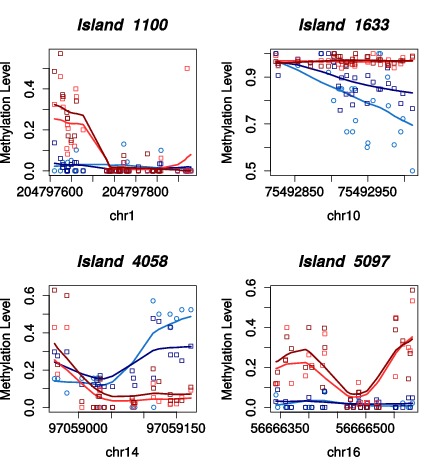}}
\caption{Methylation profiles of CpG clusters uniquely identified by the \ac{m3d} method in a comparison of leukemia (K562) and \ac{esc} cells, section 2.3 \& 3.3. \acp{dmc} are not individually very different, yet the profile has changed shape in each case.}\label{fig:01}
\end{figure}

{While these methods can be highly effective, no current method explicitly accounts for spatial covariation (MAGI implicitly assumes spatial homogeneity across a region).} DNA methylation levels are often strongly spatially correlated: accounting for such correlations in a testing procedure could then lead to considerable increases in statistical power. Some examples of spatially correlated changes in the ENCODE data analysed in Section \ref{humanData} are shown in Figure \ref{fig:01} {and Supplementary Figure S1}; notice that in all of these examples the change at individual cytosines is modest, and hence these regions would not be called as \acp{dmr} by currently existing methods. We remark that, while local smoothing methods like BSmooth (\citealp{hansen2012bsmooth}) attempt to capture spatial coherence, the local coherence is not an integral part of the testing procedure. {Smoothing in this setting serves the dual purposes of filtering noise and highlighting large scale changes in the methylation profile}.  Moreover, the shape of the methylation profile has been suggested as an important factor in predicting gene expression (\citealp{vanderkraats2013discovering}), leading to a potentially functional role for methylation patterns. To our knowledge, there are no methods that test higher order properties, such as shape, of the methylation profiles over a region. 

Here we present \ac{m3d}, a non-parametric statistical test for {identifying \acp{dmr} from pre-defined regions}, explicitly accounting for shape changes in methylation profiles. Our method is based on the \ac{mmd}, a recent technique from the machine learning literature which tests whether two samples have been generated from the same probability distribution (\citealp{gretton2007kernel, gretton2012kernel}). Similar non-parametric tests have already been applied to ChIP-Seq and RNA-Seq data (\citealp{schweikert2013mmdiff,drewe2013accurate}). {Our contribution is to adapt the method for the specific challenges of bisulfite sequencing data, introducing an explicit control for confounding changes in coverage levels. Our method is used to test for changes in methylation profiles across regions, as opposed to individual cytosines, and we call as \acp{dmr} those regions whose variation cannot be explained by inter-replicate variability.} We demonstrate the performance of \ac{m3d} against existing methods on real and simulated data, showing a considerable increase in power and improved robustness against reduced replication and coverage levels.

\section{Methods}

The \ac{m3d} method is designed to analyse aligned methylation data. Rather than testing individual cytosines and pooling them into putative \acp{dmr}, \ac{m3d} considers changes in the methylation profile's shape over a given region. To quantify shape changes, we compute the \ac{mmd} over each region and adjust it to account for changes in the coverage profile across samples. Finally, we use a data-driven approach to compare test statistics based on the empirical likelihood of observing between-group differences among replicates. We restrict our analysis to CpGs only and combine data from both strands. 

{Selecting which regions to test is an important feature of a differential methylation study, and must reflect the specific question being asked.} Regions can either be pre-defined, such as a list of promoter regions, or generated from the data by selecting regions of dense CpGs (clusters) as in [\citealp{hebestreit2013detection}]. {We keep this as a flexible option and instead focus on a general framework for region-based methylation analysis.}

\subsection{Maximum Mean Discrepancy}

Formally, the \ac{mmd} is defined as follows. Let $\mathcal{•}al{F}$ be a class of functions $f:\mathcal{X}\rightarrow \mathbb{R}$ over a metric space $\mathcal{X}$ with Borel probability measures $p,q$. We define the \ac{mmd} as

\begin{equation}
MMD[\mathcal{F},p,q]=\sup_{f\in \mathcal{F}}{(\mathbf{E_{p}}[f(x)]-\mathbf{E_{q}}[f(x)])} \label{eq:01}
\end{equation}
Intuitively, we are finding the mean over a bounded function that maximises the difference between the probability distributions. For a sufficiently dense function class, this is equal to 0 if, and only if, $p=q$. Choosing $\mathcal{F}$ to be the unit ball in a \ac{rkhs} on $\mathcal{X}$ provides a searchable class of functions that retains this result (\citealp{gretton2007kernel}). For $x,x'$ independent random variables with distribution $p$ and $y,y'$ independent random variables with distribution $q$, the square of the MMD becomes:

\begin{equation}
MMD^2[\mathcal{F},p,q]=\mathbf{E_{p}}[k(x,x')]-2\mathbf{E_{p,q}}[k(x,y)]+ \mathbf{E_{q}}[k(y,y')] \label{eq:02}
\end{equation}

In practice, for $X = \left \{x_{1},....,x_{m}\right \}, Y= \left \{y_1,...,y_{n}\right \}$  observations \ac{iid} from $p$ and $q$ respectively, we can compute a sample-based approximation to the MMD metric, giving rise to a feature representation in the \ac{rkhs}, as

\begin{equation}
MMD[X,Y,k]=\left[ \frac{K_{xx}}{m^2}-\frac{2K_{xy}}{mn} + \frac{K_{yy}}{n^2}\right]^\frac{1}{2} \label{eq:03}
\end{equation}

\begin{align*}
\text{where} \quad K_{xx}=\sum_{i,j=1}^{m}k(x_{i},x_{j}) &\text{,} \quad K_{xy}=\sum_{i,j=1}^{m,n}k(x_{i},y_{j})\text{,} \\ 
\text{and} \quad K_{yy}=&\sum_{i,j=1}^{n}k(y_{i},y_{j})
\end{align*}

\subsection{The \ac{m3d} statistic}
We represent a RRBS data set as a set of vectors $x_{i}$, where each $x_i$ is composed of the genomic location of a cytosine $C_{i}$, and the methylation status of that $C_{i}$ on one mapped read, $x_{i}=(C_{i},Meth_{i})$. Thus, there are as many $x_i$s in a data set as the number of mapped cytosines (within a CpG context). In order to define an \ac{mmd} between data sets, we need to define a kernel function operating on pairs of vectors $x_i$, $x_j$ in order to evaluate equation \eqref{eq:03}. A natural choice is a composite kernel given by the product of a \ac{rbf} kernel on the genomic location and a string kernel on the methylation status: 
{\begin{align}
k_{full}(x_{i},x_{j}) &= k_{RBF}(x_{i},x_{j})k_{STR}(x_{i},x_{j}) \label{eq:kernels} \\
\text{Where} \quad k_{RBF}(x_{i},x_{j}) &= exp[-(C_{i}-C_{j})^2/2\sigma^2] \\
\text{and} \quad  k_{STR}(x_{i},x_{j}) &= \begin{cases}
    1, & \text{if} \quad Meth_{i}=Meth_{j}\\
    0, & \text{otherwise}
  \end{cases}
 \end{align}}

{The RBF kernel, $k_{RBF}(x_{i},x_{j})$} retains spatial information at a scale determined by the hyper-parameter $\sigma$, which corresponds to the distance along the genome that displays methylation correlation. We model this parameter independently for each region, $R$, to reflect the local correlation structure, as $\sigma_{R}^2 = \bar{x}^2/2, \text{ for } x \in R$, a heuristic suggested in [\citealp{gretton2012kernel}]. Here $\bar{x}$ refers to the median distance of all observations in region $R$ across the data sets being compared.
\ac{mmd} distances computed using the above procedure would capture both differences in coverage profiles and differences in methylation profiles. A particular challenge of bisulfite sequencing data, and a central tenet of the \ac{rrbs} procedure (\citealp{gu2011preparation}), is that the frequency with which a cytosine site is tested (the coverage) is unrelated to the methylation status. This poses a challenge in all bisulfite sequencing analysis, as the sampling distribution becomes a confounding factor in our attempt to understand methylation. We control for changes in the coverage profile by subtracting the analogous \ac{mmd} of the coverage; the \ac{m3d} metric is then given by:

\begin{equation}
M^3D[X,Y]=MMD[X,Y,k_{full}]-MMD[X,Y,k_{RBF}] \label{eq:04}
\end{equation}

\vspace{2pt}

where $k_{full}$ and $k_{RBF}(x_{i},x_{j})$ are as described in equations 4 and 5 and the $MMD$ terms are as in equation \eqref{eq:03}. {Henceforth, we refer to the the first term, $MMD[X,Y,k_{full}]$ as the 'full MMD' and the second term, $MMD[X,Y,k_{RBF}]$, as the 'coverage MMD' for convenience.}

The last term in equation \ref{eq:04} represents the \ac{mmd} of the data on a methylation-blind subspace. This implies that, in the large sample limit when the sample estimate of the \ac{mmd} converges to the exact \ac{mmd} of equation \eqref{eq:01}, the \ac{m3d} statistic is non-negative. 

The \ac{m3d} statistic will therefore be different from zero when there is a change in the methylation profile, independently of a change in the coverage profile. As a consequence, \ac{m3d} between replicate \ac{rrbs} experiments (which do not necessarily have identical coverage) should be close to zero or, equivalently, the full \ac{mmd} should be equal to the {coverage} \ac{mmd}. This is borne out in the data; the metrics strongly agree over replicates. Testing equality of metrics over 102 ENCODE RRBS data sets gives an $R^2$ of $0.95$. This can be seen in Supplementary Figure S2; specific examples can also be seen in Figure \ref{fig:02} (a-c) and Figure \ref{fig:03}(a-c), where the dense region around the diagonal represents unchanged \acp{dmr} with \ac{m3d} close to zero. 

\subsection{$P$-value Calculation}

We use the \ac{m3d} as a test-statistic {by comparing the values observed across groups to those observed between replicates. We define our null distribution as the observed \ac{m3d} values over all of the testing regions between all replicate pairs. For a given region $r$, we compute the mean, $\mu_{r}$ of the \ac{m3d} values over all sample pairs across testing groups for $r$. The $p$-value for $r$ is the probability of observing $\mu_{r}$ or higher among the null distribution.} We use the Benjamini-Hochberg procedure to calculate \acp{fdr}, rejecting clusters at a 1\% significance level (\citealp{benjamini1995controlling}). Since each test corresponds to an entire region, this correction is less punitive than methods testing each cytosine location. 

{In general, we calculate the $p$-value empirically. In order for the method to scale, we also provide a model based approximation by fitting an exponential distribution to the 95th percentile of the null distribution. $P$-values are calculated in the same manner using the fitted exponential. An example is shown in Supplementary Figure S3.}

{At a given FDR cut-off, identifying \acp{dmr} amounts to identifying a threshold \ac{m3d} value, $t_{fdr}$ and calling all regions $r$ with $\mu_{r}>t_{fdr}$. We find the empirical method to be marginally more accurate (see Figure \ref{fig:ROC_sim}) and report results with empirically calculated $p$-values for the rest of this paper.}

\section{Data sets}
We benchmarked the \ac{m3d} method on a simulated data set and two real data sets. We briefly describe here the two real data sets as well as the simulation procedure used. {Summary statistics and figures of the testing regions are shown in section S4 of the Supplementary Data.}

\subsection{Simulated data} \label{simulatedData}
To benchmark the ability of our method to detect true changes without introducing false positives, we resort to a simulation study. In order to simulate methylation profile changes with realistic statistics, we constructed our simulation from a real \ac{rrbs} data set. 

{We used an RRBS data set of human embryonic stem cells, H1-hESC (described in the following section), consisting of 2 replicates. Dense CpG regions were identified using the procedure in (\citealp{hebestreit2013detection}), and for simplicity we focused on the first 1000 on chromosome 1. We then simulated two more replicates to act as our testing group, as described in section S5 of the Supplementary Data. The coverage statistics for the resulting data set have a mean of 34.5, and a median coverage of 23 at each CpG site. We then selectively altered the methylation profile of randomly chosen regions in the simulated replicates to create {known} methylation changes and used the \ac{m3d} method to test for \acp{dmr}.}

To simulate methylation changes, we randomly selected 250 of the CpG clusters out of a possible 1000. We selected a short region within each cluster, at least 100bp long and with a total coverage of at least 100, in each replicate. {If necessary, we increased the size of the region until it occupied at least $n$ CpG sites, where $n$ was uniformly sampled from $[4,20]$.} The methylation level, $L^{old}_{i}$, was calculated at each cytosine site, $C_{i}$ as the proportion of all the data points mapping to that site that were methylated. We measured the mean methylation of the sites and created a simulated methylation level, $L^{new}_{i}$, by hyper-methylating the region if it was less than 50\% methylated on average, and hypo-methylating it otherwise. The degree of methylation change was controlled by a parameter $\alpha \in [0,1]$, such that the new methylation level $L^{new}_{i}=(1-\alpha)L^{old}_{i} + \alpha$ if the region was being hyper-methylated and  $L^{new}_{i}=(1-\alpha)L^{old}_{i}$ if it was being hypo-methylated. To vary the strength of methylation change, we tested the methods different values of alpha.

Simulated data was then created by sampling data points $\left \{ x_{1},...,x_{n_{i}} \right \}$ at site  $C_{i}$ with corresponding $\left \{ Meth_{1},...,Meth_{n_{i}} \right \}$ sampled with probability $p(Meth_{j}=methylated)=L^{new}_{i}$, where $n_{i}$ is the coverage at location $C_{i}$. Pseudocode for creating simulated data is shown in Section S5 of the Supplementary Data.

\subsection{Human Data}\label{humanData}

To test the \ac{m3d} method on real data, we compared two Tier 1 tracks from the ENCODE consortium, GEO series GSE27584 (\citealp{encode2012integrated}). RRBS data from human embryonic stem cells, H1-hESC, were compared against leukemia cells, K562. Both data sets were produced by the Myers Lab at the HudsonAlpha Institute for Biotechnology. The data is available pre-processed and aligned to the hg19 genome, and we used the resulting BED files. H1-hESC cells came from a human male and K562 from a female, so sex chromosomes were removed from the analysis. Testing regions were defined by clustering CpG sites in the same manner as for the simulations and regions with no coverage in at least one sample were excluded from the analysis; this resulted in 14,104 genomic regions for testing.

To investigate the relationship between differential methylation and gene expression, we used the corresponding 200bp paired end RNA-seq data for H1-ESC and K562 cells available in release 4 of the ENCODE consortium (\citealp{encode2012integrated}). Reads were aligned using TopHat and gene expression estimates in \ac{fpkm} were produced with Cufflinks (\citealp{trapnell2012differential}). Gene expression estimates were averaged across the 3 replicates within each group and analysis was performed on the resulting changes across groups.

\subsection{Mouse Data}\label{mouseData}

We compared a 4 replicate data RRBS data set from mouse strain B6C \acp{esc} (GEO: GSE56572, \citealp{booth2014quantitative}) to a 3 replicate data set consisting from sciatic nerve cells from postnatal day 10 (P10) mice (GEO: GSE45343, \citealp{varela2014s}). To define testing regions, we used the list of exons for Mus musculus provided by Ensembl in release 75 (\citealp{flicek2013ensembl}). We excluded any exon regions with no coverage in one of the ESC cell samples or with less than 5 CpG sites in total, leaving 2359 regions in total. Again, data from both strands were combined. The median coverage at the remaining CpG loci was 24 for the \acp{esc} and 11 for the P10 sciatic nerve cells.

\section{Results}

\subsection{Simulations}

\begin{table}[t]{Table 1: Simulation Results. Sensitivity to low Methylation Changes \vspace{0.3cm}}\label{Tab:01}
\begin{adjustwidth}{-2cm}{}
{\begin{tabular}{l|| c c c| c c c| c c c| c c c|}\hline
Alpha  & \multicolumn{3}{c|}{1} & \multicolumn{3}{c|}{0.8}& \multicolumn{3}{c|}{0.6} & \multicolumn{3}{c|}{0.4} \\
Meth Change (Mean, SD) & \multicolumn{3}{c|}{(0.96, 0.10)} & \multicolumn{3}{c|}{(0.76,0.15)}& \multicolumn{3}{c|}{(0.57,0.16)} & \multicolumn{3}{c|}{(0.38,0.15)}\\ \hline
 & \ac{m3d} & BS & {MAGI} & \ac{m3d} & BS & {MAGI} & \ac{m3d} & BS & {MAGI} & \ac{m3d} & BS & {MAGI} \\ \hline
Correct & 232 & 67 & {211} & 231 & 65 & {205} & 210 & 66 & {201} & 197 & 66 & {157}\\
Type-1 & 0 & 10  & {2} & 0  & 7 &{2} & 0 & 10 & {2} & 0 & 10 & {0}\\ 
Type-2 & 18 & 183 & {39} & 19 & 185 & {45} & 40 & 184 &{49} & 53 & 184 & {93} \\ \hline
\end{tabular}}
{\\ \vspace{0.3cm} For various values of alpha, we show the corresponding mean and standard deviation of the methylation level change (the total methylated reads divided by coverage at each CpG) for the altered CpG sites and the results of testing the three methods. MMD outperforms BSmooth and MAGI \vspace{0.5cm}.}
\end{adjustwidth}
\end{table}
\vspace{0.3cm}

\begin{figure}[!tpb]
\centering
\includegraphics[width=0.7\textwidth]{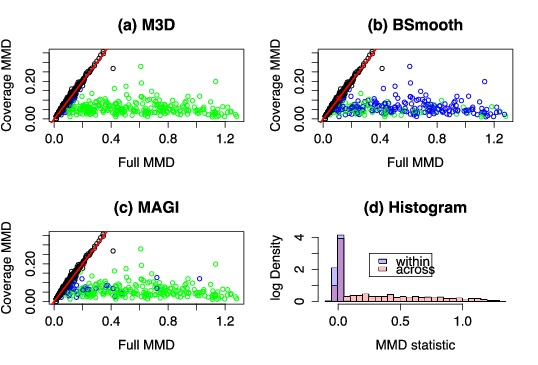}
\caption{Simulation Results. (a-c) We plot here the {coverage} MMD against the full MMD metric {for all methods. The \ac{m3d} test statistic is their difference}, the distance in the x-axis from the red line. Each point is a CpG cluster. Black are unchanged, Green are correctly called \acp{dmr}, Blue are missed \acp{dmr}, Red are incorrectly called clusters. (a) \ac{m3d} identifies a clear relationship and calls almost all of the clusters. (b) BSmooth calls some of the clusters but makes both types of error (see Table 1). Classification bears little resemblance to the \ac{m3d} method. (c) {MAGI calls fewer regions, again with little semblance to the \ac{m3d} method}  (d) Histogram of test statistics for replicate values (blue) and with simulated changes (red), log scale.} \label{fig:02}
\end{figure}

We first benchmarked our method on a realistic simulated data set generated as described in Section \ref{simulatedData}. Results were compared against {BSmooth and MAGI}. BiSeq and methylSig were also considered; however, since the data set had lower replication than the minimum recommended by the authors, we decided not to use them. BSmooth was designed for \ac{wgbs} data; in order to adapt the method to \ac{rrbs} data, we followed the authors' suggestion and altered the maximum allowable distance between neighbouring cytosines before smoothing\footnote{bioconductor mailing list: https://stat.ethz.ch/pipermail/bioconductor/2013-February/051020.html}. {Details for implementation of BSMooth and MAGI are provided in section S6 of the Supplementary Data.}

\begin{figure}[!tpb]
\centering
\includegraphics[width=0.4\textwidth]{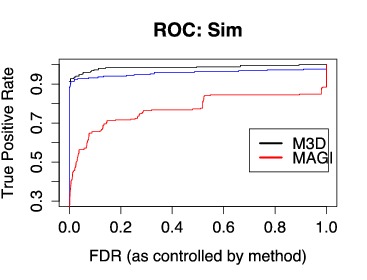}
\caption{{ROC curve. Here we plot the true positive rate against the FDR for each method, reflecting the proportion of regions called at each FDR. Black is \ac{m3d}, with empirical $p$-values, Blue is \ac{m3d} with modelled $p$-values and red is MAGI. AUCs for the methods are 0.99, 0.96 and 0.77 respectively }} \label{fig:ROC_sim}
\end{figure}

Figure 2 summarises the results obtained with the methylation strength parameter $\alpha$ (see Section \ref{simulatedData}) set to 1. Of the 250 differently methylated regions, the \ac{m3d} method called 232, with no falsely called \acp{dmr}. Figures \ref{fig:02}a-c show scatterplots of coverage \ac{mmd} on the $y$ axis vs full \ac{mmd} on the $x$ axis for all 1000 regions, with colours denoting the results of the testing procedure using the different statistics.  Individual regions are represented as circles, coloured according to whether the region was a true positive (green), a false positive (red), a false negative (blue) or a true negative (black). As discussed before, changes in methylation are likely to occur for regions that are mapped far from the diagonal. The figures show a clear cluster of regions about the diagonal (the unchanged regions) and a clearly identifiable group with much larger full \ac{mmd} (the changed regions). Figure \ref{fig:02}a shows the results of the testing procedure using the \ac{m3d} statistic. As we see, \ac{m3d} correctly identifies most of the {250 simulated} changes. {ROC curves are shown in Figure \ref{fig:ROC_sim}. Note that BSmooth is omitted as the method does not test regions as a whole, rather identifies groups of \acp{dmc} within the region, and hence does not output a relevant statistic for comparison.}

{We present the results of BSmooth {and MAGI} in same framework in \ref{fig:02}b-c.} BSmooth correctly called 67 of the regions with an additional 10 false positives, typically calling regions with similar coverage profile, which we expect is due to the effect of local likelihood smoothing. Figure \ref{fig:02}b shows the BSmooth results; as we see, even regions with very marked shape differences (as quantified by \ac{m3d}) were missed, which is to be expected as BSmooth does not include spatial correlations in the testing procedure. {MAGI called 211 of the regions correctly, with 2 false positives (the FDR was set to 1\%). Figure \ref{fig:02}c shows that while many of the regions missed had a low \ac{m3d} statistic, this was not always the case and there is not a simple relationship between the two methods indicating that the \ac{m3d} method performs a genuinely different computation, as opposed to simply being more powerful}. A histogram of the \ac{m3d} test statistic is shown in Figure \ref{fig:02}d for the replicates and the cross-group comparisons. The empirical testing distribution is shown in blue and is seen to be consistent around zero and sharply peaked. The simulated \acp{dmr} are easily distinguished.

We then investigated the sensitivity of our method by systematically altering the strength of the methylation changes using $\alpha = 1,0.8,0.6 \text{ and } 0.4$. We compared the \ac{m3d} method to {BSmooth and MAGI} and show our results in Table 2. The \ac{m3d} method maintains a very creditable performance level for {all of the $\alpha$ values. This is due to the fact that neighbouring cytosines are being altered, and hence there are spatial correlations in the changes. Were the changes scattered randomly in the region \ac{m3d} performance would weaken, whereas MAGI would remain robust. The sudden dip in performance of MAGI at $\alpha=0.4$ is due to fewer of the cytosines' methylation levels crossing the threshold value}. It is remarkable that at all levels of $\alpha$ the use of the \ac{m3d} statistic does not lead to any type I errors, {though we note that the other methods have consistent type 2 error levels across these tests.}

{To assess the sensitivity of the various methods to spatial correlations, we ran a further simulation, this time adding a 'Gaussian bump' of to the methylation profiles of the regions, at randomly chosen locations, with varying widths and strengths. Details are described in Section S5 of the Supplementary Data. Here we found a more marked contrast in the performance of the methods, which we show in Table 2. ROC curves are shown in Section S7 of the Supplementary Data.}\\

\begin{table}[!t]
\centering
{Table 2: Gaussian Bump Simulation\\ \vspace{0.3cm}}\label{Tab:02}
{\begin{tabular}{l|| c c c|}\hline
 & \ac{m3d} & BS & {MAGI}  \\ \hline
Correct & 190 & 58 & {102} \\
Type-1 & 0 & 11  & {1} \\ 
Type-2 & 60 & 192 & {148}  \\ \hline
\end{tabular}}{\\ \vspace{0.3cm} Results for adding a Gaussian bump to the methylation levels. Despite the overall change being smaller, \ac{m3d} retains good performance by considering spatial correlations in the data.}
\end{table}

\subsection{Human Data}

\begin{figure}[!tpb]
\centerline{\includegraphics[width=0.48\textwidth]{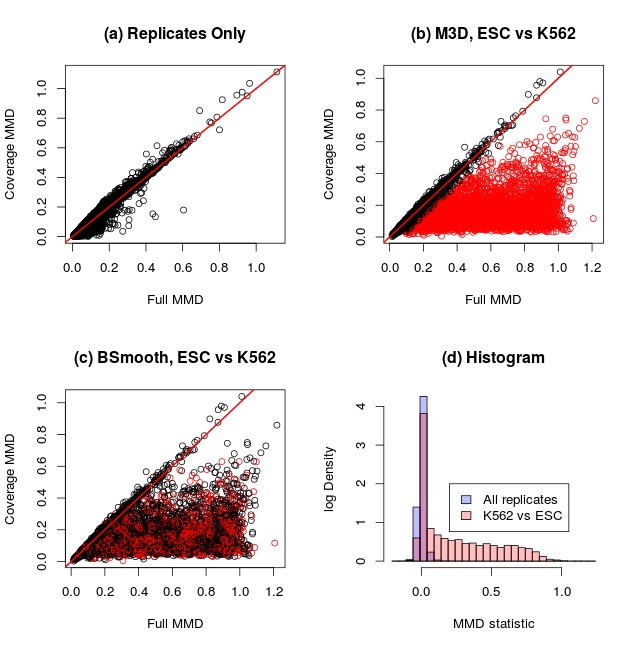}}
\caption{H1-hESC vs K562 Cells. Black dots are uncalled clusters, red are called. (a) Just the inter-replicate metrics are shown, for comparison with Figure \ref{fig:02}. (b) Between-group clusters as called by \ac{m3d}. (c) BSmooth identifies far fewer. {Axes show the full and coverage MMD.} Classification bears little resemblance to the \ac{m3d} method. (d) The histogram of test statistics.} \label{fig:03}
\end{figure}
We now describe the results of comparing two human data sets generated by the ENCODE consortium (see Section \ref{humanData}). We focus on three aspects: a general comparison of results between \ac{m3d} and BSmooth, an analysis of the robustness of the results with respect to the coverage, and an analysis of the functional relevance of our results.

\subsection{DMR detection.} Out of the 14,104 CpG regions selected for testing (see Section \ref{humanData}), \ac{m3d} identified 4137 \acp{dmr}, {and BSmooth and MAGI identified 1649 and 3101 \acp{dmr}, agreeing on 1328 and 2353 of the regions respectively. In Figure \ref{fig:03} we present the results in the same form as Figure \ref{fig:02}, where we again see that the \ac{m3d} method produces genuinely different results.} Figure \ref{fig:03}a shows the method applied to the replicates only, where the methylation-blind and aware metrics agree. Figures \ref{fig:03}b and \ref{fig:03}c show the results of between-group testing by \ac{m3d} and BSmooth respectively. The data has a striking similarity to Figure \ref{fig:02}a; this suggests that, on this real data set, the \ac{m3d} statistic provides an excellent measure of changes in methylation profiles. Similarly to the simulated data set, BSmooth does not behave in a consistent manner with regard to the \ac{m3d} test-statistic and many CpG clusters are missed (Figure \ref{fig:03}c). A histogram of the \ac{m3d} test statistics is shown in Figure \ref{fig:03}d, again confirming that the \ac{m3d} statistics identifies a clear group of changed profiles between the two conditions. {Comparisons to MAGI are shown in section S8 of the Supplementary Data.}

\subsubsection{Robustness to Low Coverage}

To test the consistency of the \ac{m3d} method to low coverage, we simulated a reduction in the coverage levels of the H1-hESC and K562 data. We discarded at random reads for the data sets to simulate a reduction in coverage by 75\%, 50\% and 25\% in both data sets. The \ac{m3d} method was used to find \acp{dmr} and the results were compared across coverage levels; to alleviate the computational burden, we only considered CpG regions on chromosome 1.

\begin{figure}[!tpb]
\centerline{\includegraphics[width=0.3\textwidth]{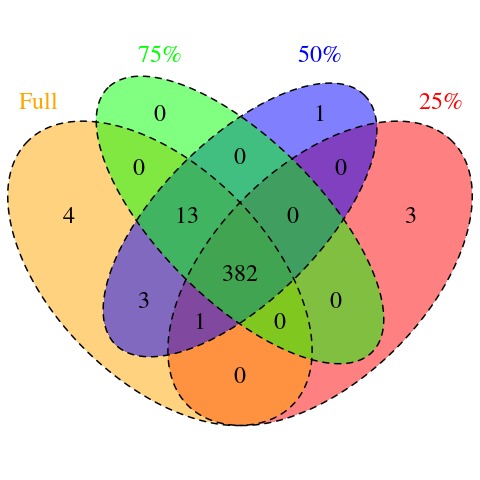}}
\caption{Venn Diagram of Calls with Reduced Coverage. 382 calls are consistent at all coverage levels. The method misses clusters at lower coverage levels, yet it does not {call many DMRs that weren't identified at higher coverage levels.}} \label{fig:04}
\end{figure}

Of the 1345 CpG clusters on chromosome 1, the \ac{m3d} method identified 403 \acp{dmr}. Reducing coverage to 75\%, 50\% and 25\% of the original level, the method identified 395, 399 and 386 \acp{dmr} respectively, with 0, 1 and 1 \acp{dmr} not in the original set. To see the consistency of the calls, we show a Venn Diagram in Figure \ref{fig:04}. There is a strong overlap in the calls at all levels, indicating robustness in the method. Interestingly, while fewer regions are called at lower coverage levels, at each stage we do not see many new regions being called as might be expected. While the empirical testing distribution is less accurate, the structure remains intact. Both BSmooth and MAGI showed similar consistency, although the number of regions called was lower in both cases. Results are shown in section S9 of the Supplementary Data.

\subsubsection{Functional analysis of differential methylation regions.}

To investigate the functional relevance of our results, we interpreted the called \acp{dmr} in terms of the functional annotation and expression level of nearby genes. We used three sets of genomic regions, gene bodies, first exons and promoters;
gene and exon regions were downloaded from  Ensembl release 75 (\citealp{flicek2013ensembl}) and promoter regions were defined as being 2000bp upstream from the \ac{tss}. Tests were run on gene, promoter and first exon regions separately. We excluded regions with fewer than 5 CpGs and tested the remaining regions for changes in the methylation profiles using the \ac{m3d} method.

Since the \ac{m3d} method provides a measure of the strength, but not direction of the methylation profile change, we measured the cross group expression change as the absolute value of the \ac{fpkm} log-fold change and excluded all genes with less than 100 \ac{fpkm} in one sample to avoid noise at low expression levels. 

We identified 8747 gene body regions with sufficient expression and CpG content; among these, the \ac{m3d} method called 404 as \acp{dmr}. We tested the absolute log-fold changes in expression between called and uncalled regions with a Wilcoxon Rank-Sum test and found the former was higher (p-value: $2.2\times10^{-16}$). Similarly, 103 of 4916 promoter regions were called as \acp{dmr} and showed an associated higher absolute log-fold expression change (Wilcoxon Rank-Sum test, p-value: $1.18\times 10^{-6}$). Rather more first exon regions were tested, with 411 of 19473 regions being called as \acp{dmr}. Again, there was an associated increase in the log-fold expression change (Wilcoxon Rank-Sum test, p-value: $2.2\times 10^{-16}$). None of the 411 first exon \acp{dmr} were in the gene bodies of the gene region testing group, hence this is not a duplicate result.
The median log-fold expression change for genes associated with uncalled regions was $0.15$ for gene, promoter and first exon regions. In genes associated with called promoter regions this rose to $0.24$, and in genes and first exon regions the median was $0.37$ and $0.34$ respectively (Supplementary Figure S12). These results support earlier studies outlining a stronger link between methylation in the first exon, as opposed the promoter region, and gene expression (\citealp{brenet2011dna}).

We performed an enrichment analysis for \ac{go} terms using the Ontologizer software for the gene, promoter and first exon regions separately (\citealp{bauer2008ontologizer}). In each case, the population group was chosen to be the set of genes associated with the regions being tested, i.e. those with sufficient coverage and CpG counts, and the study group was the set of genes associated with the called regions. For this study, we examined the \acp{dmr} called by the \ac{m3d} method against all of the regions tested, independently of gene expression data. We used parent-child analysis and adjusted p-values according with a Benjamini-Hochberg procedure at 10\% \ac{fdr}.

For the gene regions, we tested 2,692 called genes against 15,321 tested genes resulting in 208 \ac{go} terms being called. Among these, \ac{go} terms for embryonic morphogenesis , organ formation, growth, developmental growth, regulation of cell differentiation, pattern specification process and cell fate commitment were discovered, as might be expected in a comparison between \acp{esc} and fully mature K562 cells. This suggests a connection between gene body methylation and cell function.

The first exon group was smaller, with 1,087 called gene associations in a population of 10,811. 31 \ac{go} terms were statistically significantly enriched. 24 of these terms were also enriched in the gene region analysis. Again, terms associated with cell differentiation, such as cell fate specification and cell fate commitment, were observed. This is striking, since these first exons were not in gene bodies of the gene testing regions.

The promoter group had 506 gene associations called out of a population of 8,114. Interestingly, we found no statistically significantly enriched \ac{go} terms in the analysis. The top ten enriched \ac{go} terms, by statistical significance are shown in Supplementary Tables 1, 2 and 3 in section 11 for the three types of genomic regions tested.

\subsection{Mouse Data}
In order to examine the robustness of the \ac{m3d} statistic to changes in replication, we considered a comparison between two mouse data sets with larger replication, the \acp{esc} data set of (\citealp{booth2014quantitative}) with four replicates, and the neural data set of (\citealp{varela2014s}) with three replicates. Robustness to low replication is important; as remarked before, while many methods require at least three replicates in each data set, many experimental protocols (including almost all of the ENCODE \ac{rrbs} data) provide only 2 replicates. We used the \ac{m3d} method to identify \acp{dmr} with 3 and with 2 ESC replicates, and compared the set to those identified with the full 4 ESC replicate sets. \acp{dmr} were identified  at a 1\% \ac{fdr}.

\begin{figure}[!tpb]
\centerline{\includegraphics[width=0.25\textwidth]{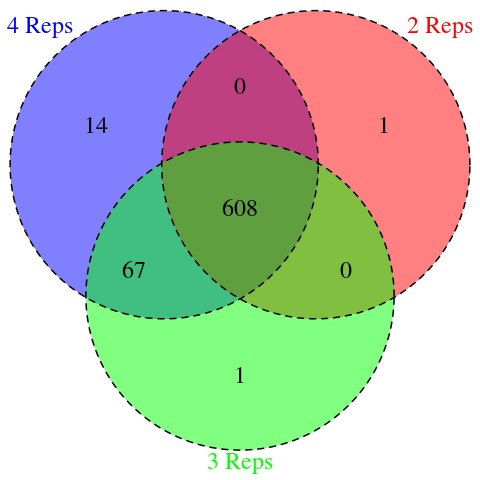}}
\caption{Venn Diagram of Calls with Fewer Replicates, for the case of 4, 3 and 2 replicates for ESC cell control group.} \label{fig:RobustReps}
\end{figure}

Of the 2359 exon regions tested, the \acp{m3d} method identified 689, 676 and 609 with methylation profiles that differed significantly with respect to inter-replicate variation with 4, 3 and 2 replicates in the ESC group respectively. As is shown in Figure \ref{fig:RobustReps}, the overlap between the three sets of called regions accounts for almost 90\% of the total. Importantly, while the testing lost power with lower replication (as can be expected), only one additional region was called, indicating that the method does not introduce many false positives with reduced replication levels.

\subsection{Computing Times}

{We report the running times for non-parallel implementations of all methods on the data set used in sections 3.2 and 4.3. This data set represents a complete RRBS experiment consisting of 14,104 testing regions. On an ordinary desktop PC, MAGI took approximately 30 seconds, BSmooth took almost 6 minutes (including smoothing) and \ac{m3d} took approximately 2 hours. The \ac{m3d} algorithm is linear in the number of testing regions and combinatorial in the number of replicates.}

%
%

\section{Discussion}

We proposed the first kernel-based test for \acp{dmr} which exploits higher order spatial features of methylation profiles. Empirical comparisons on simulated and real data show a considerable increase in statistical power in comparison with the widely used BSmooth method (\citealp{hansen2012bsmooth}), as well as considerable robustness to low coverage and low replication. {The \ac{m3d} method also outperforms MAGI (\citealp{baumann2014magi}) in our simulations, as well as calling more \acp{dmr} in the real data set, though this comes at a computational cost}.  

The increased power of the \ac{m3d} approach is due to a number of factors. Firstly, the method is sensitive to spatially correlated changes in methylation profiles. Methylation profiles are known to be highly spatially correlated in general, and the results of our experiments imply that spatial correlation is also a feature of differences in methylation profiles between conditions, at least in the data sets considered.  Secondly, the method explicitly accounts for differences in the coverage profiles between conditions, a confounding factor for other methods, as demonstrated in Figure \ref{fig:02}. Thirdly, the method models inter-replicate variability on a regional basis along the whole genome. Each regional cross-group methylation change is compared to this distribution, and test statistics for each region represent how well the change in methylation profiles can be explained by inter-replicate variability. At present, other methods that consider inter-replicate variability do so on a CpG site-by-site basis, which lacks power with low replication and coverage and does not consider regional, spatial changes. When testing the method with different strengths of methylation change at CpG loci, we saw a sharp decrease in the number of regions being called as the methylation profile change over the regions became comparable to inter-replicate variability. Other methods experienced a less pronounced change in this regard.

Other studies have suggested that changes in shape of methylation profiles are important in predicting gene expression (\citealp{vanderkraats2013discovering}). To test whether our method is able to capture functionally important changes in methylation profiles, we performed gene expression analysis with human data and showed a link between methylation changes called by the \ac{m3d} method and gene expression changes between conditions. Further, the results support the hypothesis that gene expression is more closely linked to methylation in the first exon of a gene than to methylation in promoter regions (\citealp{brenet2011dna}). \ac{go} analysis of first exon and whole gene methylation changes both revealed links to cell function, despite none of the exons overlapping the gene regions, a result that was not apparent for promoter methylation changes.  While our findings confirm a potential role for methylation profile shape as a predictor of gene expression, they do not provide biological mechanisms for linking methylation shape to gene expression and regulation. {While sequence variants and protein binding have been shown to be predictive of epigenetic variability \citep{gertz2011analysis,benveniste2014transcription}, we believe that further investigation of the mechanistic underpinnings of changes in methylation shape could be a valuable direction for research.}

{The \ac{m3d} method provides a considerable increase in power over existing methods, yet it comes at the cost of computational intensity. In this study we have restricted comparisons to datasets of low replication, as beta-binomial methods should prove effective with higher replication. We have also focused on sub-megabase scale changes for two reasons. Firstly, such an analysis is likely to be exploratory, in the sense that testing regions are not pre-defined, a key requirement for our method, and secondly, because BSmooth has proved adept at identifying large scale changes in this setting and is computationally cheaper.}

The \ac{m3d} framework was developed with \ac{rrbs} data in mind, yet, given its robustness to lower coverage, we expect that it may also be well suited for \ac{wgbs} data. In the future, it will be interesting to develop models that explain the predictive power of methylation profiles in terms of other epigenetic marks.

\section*{Acknowledgement}
 {We thank Prof. Rebecca Doerge for kindly sharing the MAGI code with us.}

\textbf{Funding}$\colon$ TM is funded by grants EP/F500385/1 and  BB/F529254/1 from the UK Engineering and Physical Sciences Research Council, UK Biotechnology and Biological Sciences Research Council, and the UK Medical Research Council. GSch acknowledges support from the EU FP7 Marie Curie Actions. GS is funded by the European Research
Council through grant MLCS306999.

\bibliographystyle{natbib}
\bibliography{document}

\section*{Appendix: Supplementary Data}
\addcontentsline{toc}{chapter}{First unnumbered chapter}
\setcounter{section}{0}
\renewcommand*{\theHsection}{chX.\the\value{section}}
\setcounter{figure}{0}
\setcounter{table}{0}

\renewcommand{\thesection}{S\arabic{section}}  
\renewcommand{\thetable}{S\arabic{table}}  
\renewcommand{\thefigure}{S\arabic{figure}}

\section{Further Examples of Unique Calls}

\begin{figure}
\begin{subfigure}[b]{1\textwidth}
\centering \includegraphics[width=0.6\textwidth]{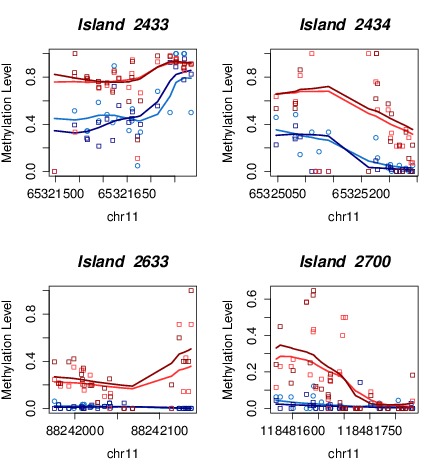}
\end{subfigure}\\
\begin{subfigure}[b]{1\textwidth}
\centering \includegraphics[width=0.6\textwidth]{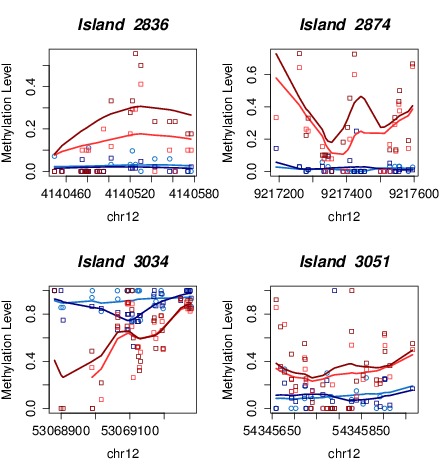}
\end{subfigure} \\
\caption{Examples of regions uniquely called by M$^3$D} \label{fig:supp_shape_examples}
\end{figure}

Further to the examples of regions uniquely called by the M$^3$D method in section 4.3 (Human Data) shown in Figure 1 in the main text, we include some more examples of shape based changes called by the method in Figure \ref{fig:supp_shape_examples}.

\section{Equality of Metrics}

\begin{figure}[!h]
\centerline{\includegraphics[width=0.5\textwidth]{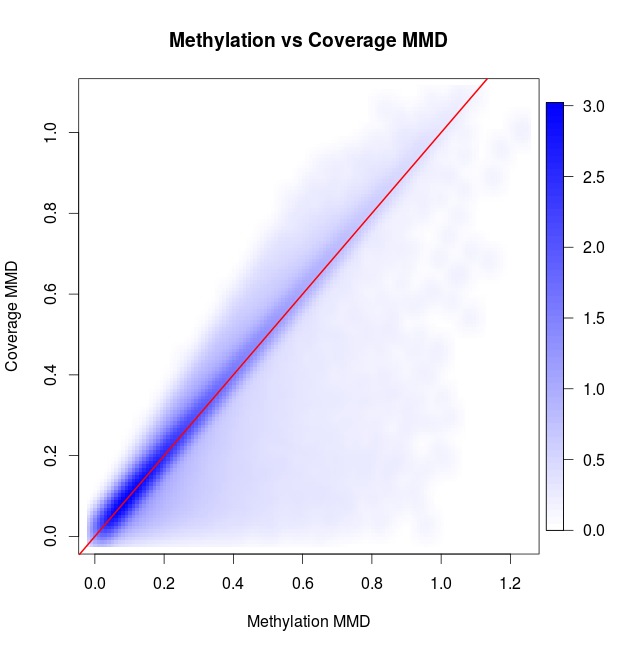}}
\caption{MMD over each region for the coverage kernel against the full kernel for 102 ENCODE RRBS data sets. There is strong agreement between the metrics.}\label{fig:supp01}
\end{figure}

Figure \ref{fig:supp01} shows a scatter-density plot of the MMD over each region for the coverage data against the full data for 102 ENCODE RRBS data sets. Testing equality of metrics gives an $R^2$ of 0.95, and we see some variation, representing inter-replicate variability.

\section{Model approach to p-values}

We fit an exponential distribution to the tail of the empirical distribution (the histogram of M$^3$D values between replicates), for the 95th percentile. This gives us a value for the exponential parameter $\lambda$, from which we calculate p-value, for a M$^3$D value of $x$ by $e^{-\lambda x}$ (which is 1 minus the cumulative distribution function). 

\begin{figure}[h!]
\centerline{\includegraphics[width=0.5\textwidth]
{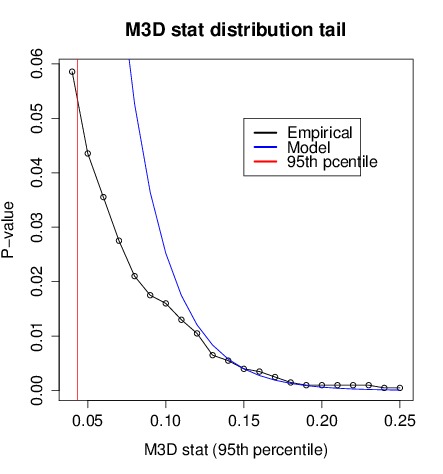}}
\caption{Modeling p-values. The black line shows the empirical distribution, the blue shows the exponential fit. The vertical red line indicates the 95th percentile. } \label{fig:fitting_exp}
\end{figure}

Figure \ref{fig:fitting_exp} shows the result of the process. We take a conservative approach and hence over estimate the p-values.

\section{Testing Regions}

\subsection{Simulation}

\begin{figure}
\begin{subfigure}[b]{0.5\textwidth}
\includegraphics[width=0.8\textwidth]{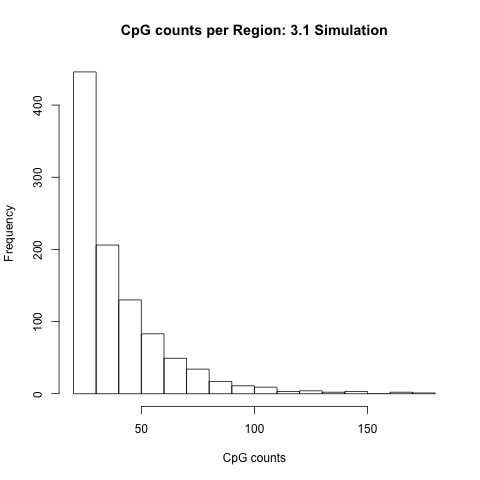}
\caption{CpG Counts}
\end{subfigure}
\begin{subfigure}[b]{0.5\textwidth}
\includegraphics[width=0.8\textwidth]{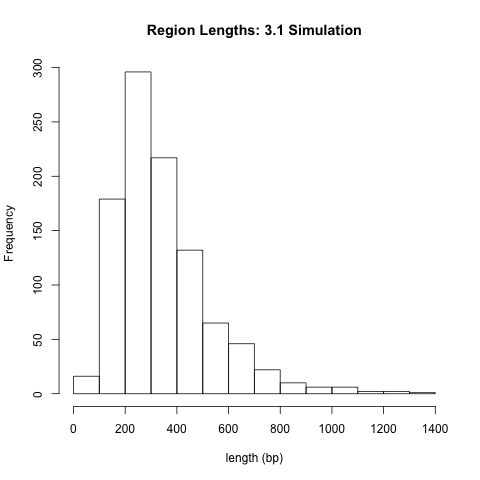}
\caption{Length of Regions}
\end{subfigure} \\

\begin{subfigure}[b]{0.5\textwidth}
\includegraphics[width=0.8\textwidth]{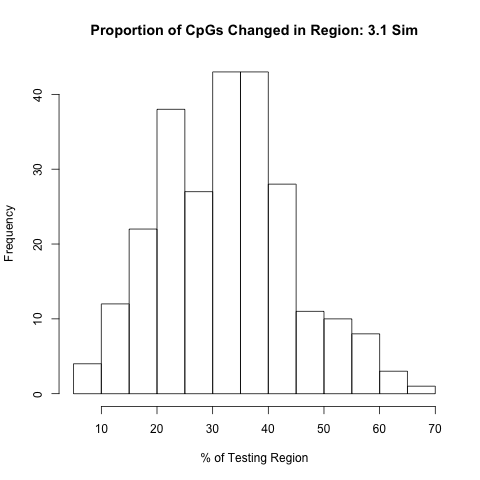}
\caption{\% CpGs changed}
\end{subfigure}
\begin{subfigure}[b]{0.5\textwidth}
\includegraphics[width=0.8\textwidth]{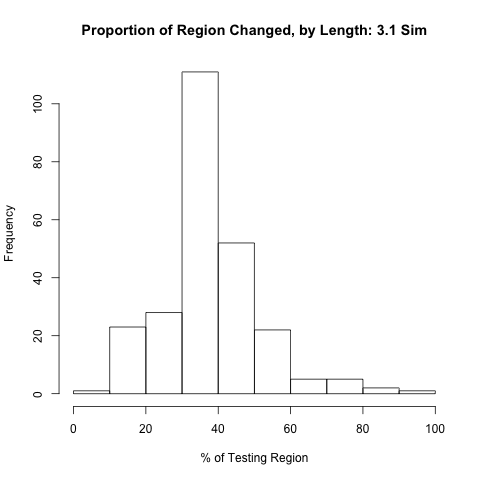}
\caption{\% region changed by length}
\end{subfigure} \\
\caption{Statistics of the testing regions for the Simulation in sections 3.1 and 4.1} \label{fig:dataSim}
\end{figure}

Figure \ref{fig:dataSim} shows histograms of the size of the testing regions for the Simulation. 

\subsection{Human Data}

\begin{figure}
\begin{subfigure}[b]{0.5\textwidth}
\includegraphics[width=0.8\textwidth]{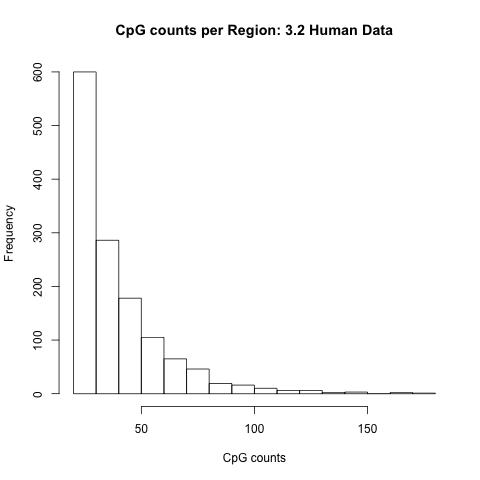}
\caption{CpG Counts}
\end{subfigure}
\begin{subfigure}[b]{0.5\textwidth}
\includegraphics[width=0.8\textwidth]{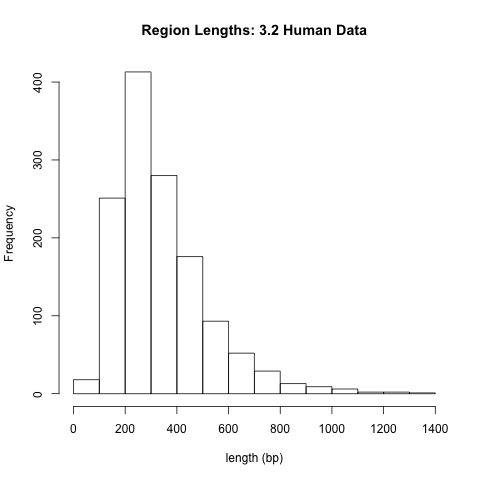}
\caption{Length of Regions}
\end{subfigure} \\

\caption{Statistics of the testing regions for the Human data in sections 3.2, 4.2}\label{fig:dataHuman}
\end{figure}

Figure \ref{fig:dataHuman} shows histograms of the size of the testing regions for the Human Data, used in sections 3.2 and 4.3. 

\subsection{Mouse Data}

\begin{figure}
\begin{subfigure}[b]{0.5\textwidth}
\includegraphics[width=0.8\textwidth]{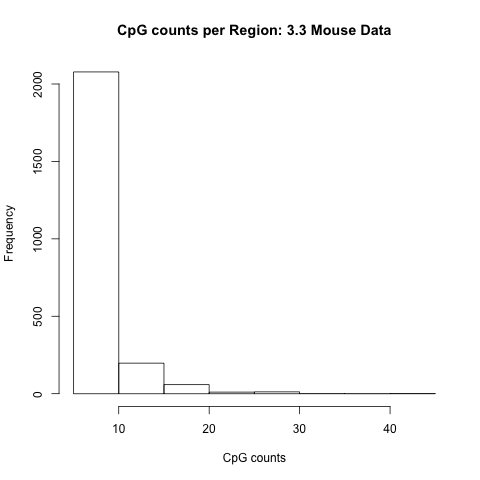}
\caption{CpG Counts}
\end{subfigure}
\begin{subfigure}[b]{0.5\textwidth}
\includegraphics[width=0.8\textwidth]{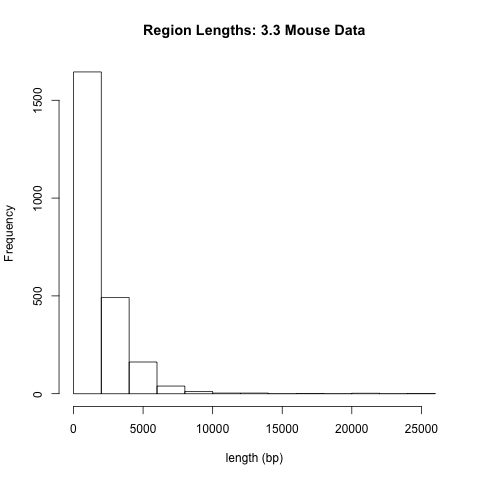}
\caption{Length of Regions}
\end{subfigure} \\

\caption{Statistics of the testing regions for the Mouse data in sections 3.3, 4.4}\label{fig:dataMouse}
\end{figure}

Figure \ref{fig:dataMouse} shows histograms of the size of the testing regions for the Mouse Data, used in section 3.3 and 4.4. 

\subsection{Testing Regions, Called vs Uncalled}

\begin{figure}
\includegraphics[width=0.8\textwidth]{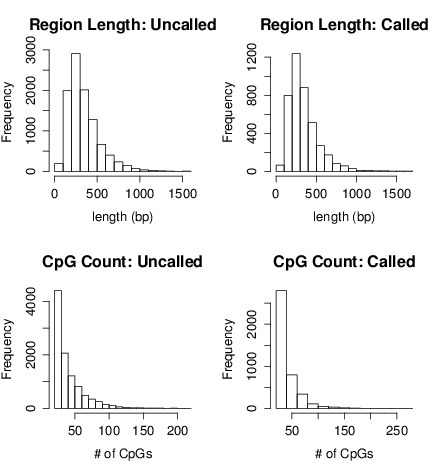}
\caption{Statistics of the testing regions uncalled and called regions, from the Human Data (section 4.3). By region length (bp) and by CpG count.}\label{fig:dataCallvsUncall}
\end{figure}

In Figure \ref{fig:dataCallvsUncall}, we show histograms of the region length and CpG count for regions that are called and not called by the M$^3$D method. The histograms in both cases have the same shape, indicating that there is unlikely to be a bias in the method for calling DMRs. The results come from section 4.3 (Human Data).

\section{Simulation Details}

The coverage profile was simulated on a region by region basis. For each region, $r$ we sampled a 'total coverage multiple', $t_{r}$ from $[0.8,1.2]$ and a 'weighting variable' $w_{r}$ from $[-0.3,1.3]$. If we call the coverage profile of the H1-hESC replicates $H1_{1}$ and $H1_{2}$ respectively, we created a coverage profile for the simulated replicate of region $r$ by taking $w_{r}*H1_{1} + (t_{r}-w_{r})*H1_{2}$. This was performed for each simulated replicate.

A simulated replicate methylation profile was created by finding the mean methylation level of the H1 replicates at each site and multiplying that by the new coverage in each simulated replicate (and rounding to ensure integer values were maintained). In the event that the H1 replicates had low coverage ($<$ 10 reads combined between them) we weighted the methylation level with a smoothed estimate of the neighbouring cytosines. The resulting coverage profiles had similar correlations to the inter-replicate values:

\begin{tabular}{l|| c c c c|}\\ \hline
 & Control1 & Control2 & Sim1 & Sim2  \\\hline
Control1 & 1 & 0.88 & 0.84 & 0.84 \\
Control2 & 0.88 & 1  & 0.87 & 0.87 \\ 
Sim1 & 0.84 & 0.87 & 1 & 0.78 \\
Sim2 & 0.84 & 0.87 & 0.78 & 1  \\\hline
\end{tabular}
\vspace{1cm}

Below we present the pseudocode describing how we simulated differential methylation in regions defined by the set \textit{truths}, for a given value of $\alpha$, which represents the strength of methylation change.

\begin{algorithm}
\caption{Change the Methylation Profile of Regions in \textit{truths}}
\begin{algorithmic}[1]
\For{\texttt{region in truths}}
	\State \texttt{Randomly select start locus}
	\While{\texttt{CpGs in section $<$ 4 OR section coverage $<$ 100 OR section length $<$ 100bp}}
		\State \texttt{Increase section by 1 locus on either side (within region)}
		\If{ \texttt{length(section) $>=$ length(region)$/3$}}
		\State \texttt{BREAK}
		\EndIf
		\EndWhile
	\For{\texttt{CpG site $i$ in section}}
		\State \texttt{$L_{i}^{old} \gets$ methylation count/coverage at i} 
		\If{ \texttt{$L_{i}^{old}<=0.5$}}
			\State \texttt{$L_{i}^{new} \gets (1 - \alpha) L_{i}^{old} + \alpha$} 
		\Else
			\State \texttt{$L_{i}^{new} \gets (1 - \alpha) L_{i}^{old}$} 		
		\EndIf
		\State \texttt{methylation count at i $\gets$ sample from $\sim B($coverage at i$,L_{i}^{new})$ } 
	\EndFor
\EndFor\\
\end{algorithmic}
\end{algorithm}

For the simulation involving adding Gaussian shaped bumps, we used the procedure in algorithm 2. $\sigma$, analogous to the variance in a normal distribution, controls the width of the change in bp,  and was drawn uniformly from the integers in the interval [4,100], while the $\alpha$ parameter, representing the strength of the change was drawn from 0.6,0.8 and 1 with equal probability.

\begin{algorithm}
\caption{Gaussian Bump: Change the Methylation Profile of Regions in \textit{truths}}
\begin{algorithmic}[1]
\For{\texttt{region in truths}}
	\State \texttt{Randomly select start locus, $l$}
	\For{\texttt{CpG site $i$ in section}}
	\State \texttt{Weight$_i$ $\gets exp[-(i-l)^2 /\sigma^2]$}  
		\State \texttt{$L_{i}^{old} \gets$ methylation count/coverage at i} 
		\If{ \texttt{Weighted mean over region of $L_{i}^{old}<=0.5$}}
			\State \texttt{$L_{i}^{new} \gets  L_{i}^{old} + \alpha \times Weight_{i}$} 
		\Else
			\State \texttt{$L_{i}^{new} \gets L_{i}^{old} - \alpha \times Weight_{i}$} 		
		\EndIf
		\State \texttt{$L_{i}^{new} \gets 0$ or $1$ if below or above resp.}
		\State \texttt{methylation count at i $\gets$ sample from $\sim B($coverage at i$,L_{i}^{new})$ } 
	\EndFor
\EndFor\\
\end{algorithmic}
\end{algorithm}

\section{Comparison Methods: Details}

BSmooth parameters were chosen to be as low as possible, whilst giving consistent performance across data sets, a minimum smoothing window (h) of 1000bp, a minimum methylation loci count (n) of 70, and a maximum gap between sites before smoothing is broken (maxGap) of 1000bp. These parameters were used for all data sets tested here.

MAGI was implemented as in the paper, using code kindly provided by the authors. The threshold value for each region was determined by k-means clustering with two clusters, as suggested in private correspondence from the authors. 


An FDR of 1\% was used for all methods, for all comparisons.

\section{ROC curves}

We present here the ROC curves for the simulation study using Gaussian bumps. We wish to determine whether pre-defined regions of interest are statistically significantly differentially methylated. MAGI takes the same approach, and hence a proper comparison is possible. BSmooth, on the other hand, is primarily a tool for discovering sections of the genome with differential methylation, relying on testing individual DMCs. The method then creates regions by chaining together CpGs with t-statistics greater than a threshold (determined from the data). This doesn't provide single, comparable statistics, and hence a direct comparison was not possible.

\begin{figure}[h!]

\includegraphics[width=0.8\textwidth]{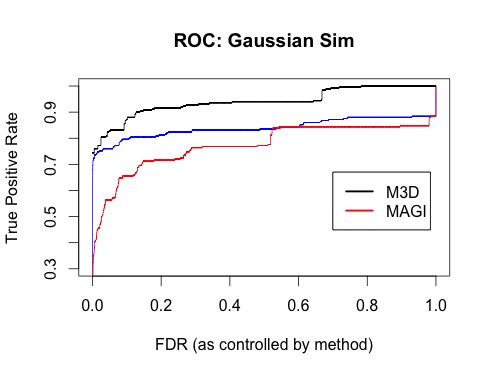}
\caption{Gaussian Sim}
\caption{ROC curve for Gaussian simulation. Here we plot ROC curves of the true positive rate against the controlled FDR for M$^3$D and MAGI. Black is the empirical p-values from M$^3$D, blue is the modelled p-values, and red is MAGI.} \label{fig:ROC_M3D_MAGI}
\end{figure}

In Figure \ref{fig:ROC_M3D_MAGI}, we show the true positive rate against the FDR, as the method assumed it. Note, this isn't the false positive rate, but the attempt to control it. This gives us a sense of the power of the method to detect DMRs at various FDRs.

We found the are under the curves for the initial simulation to be 0.99, 0.96 and 0.77 for M$^3$D empirical, M$^3$D modelled and MAGI respectively. For the Gaussian simulation, these were 0.94, 0.84 and 0.77.

\section{MAGI by M$^3$D statistics}

\begin{figure}
\begin{subfigure}[b]{0.5\textwidth}
\centering \includegraphics[width=0.5\textwidth]{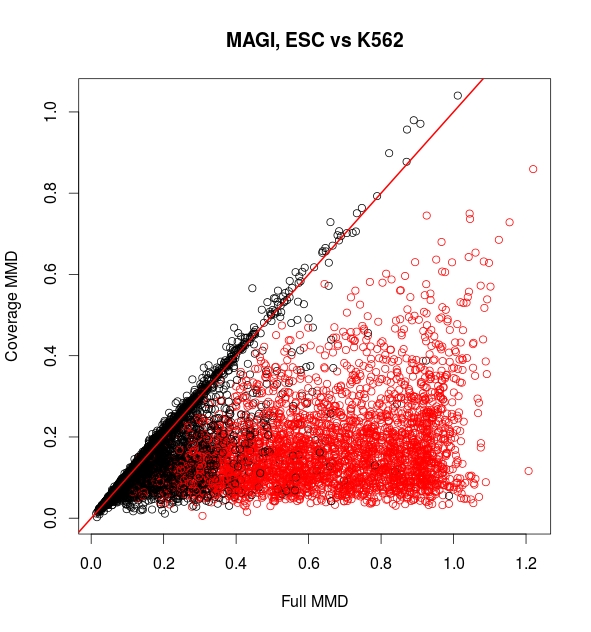}
\caption{MAGI}\label{fig:Magi_esc}
\end{subfigure}
\begin{subfigure}[b]{0.5\textwidth}
\centering \includegraphics[width=0.5\textwidth]{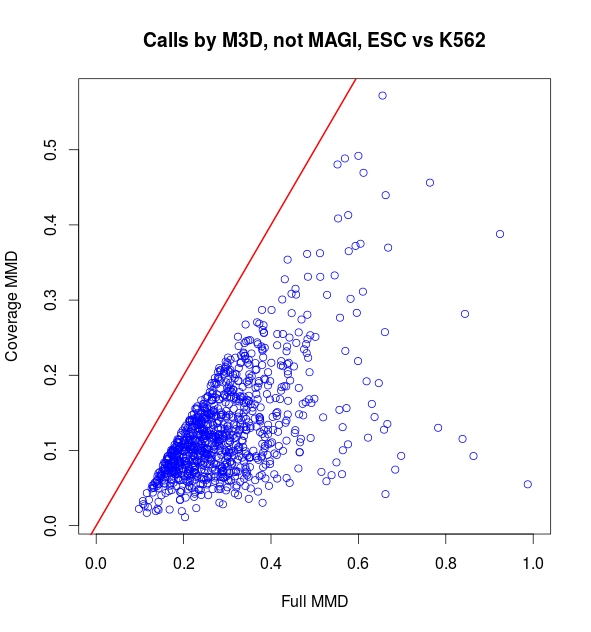}
\caption{Calls by M$^3$D, not MAGI}\label{fig:m3d_not_magi_esc}
\end{subfigure} 
\caption{MAGI, Human Data. As with Figure 3 in the main text, we plot  the coverage MMD against the full MMD metric . The M$^3$D test statistic is their difference, the distance in the x-axis from the red line. Each point is a CpG cluster. Black are unchanged, Green are correctly called DMRs, Blue are missed DMRs, Red are incorrectly called clusters. In (b) we show the regions called by M$^3$D but not MAGI.}
\end{figure}

In Figure \ref{fig:Magi_esc} we plot the calls by MAGI with respect to the M$^3$D statistics, as with Figure 3 in the main text. As noted there MAGI misses approximately a quarter of the calls M$^3$D makes. These are shown in figure \ref{fig:m3d_not_magi_esc}. MAGI calls the regions with a higher M$^3$D statistic in general, but counterexamples indicate that a fundamentally different process is at work.

\section{BSmooth \& MAGI with low Coverage}

\begin{figure}[!tpb]

\begin{subfigure}[b]{0.6\textwidth}
\includegraphics[width=0.5\textwidth]{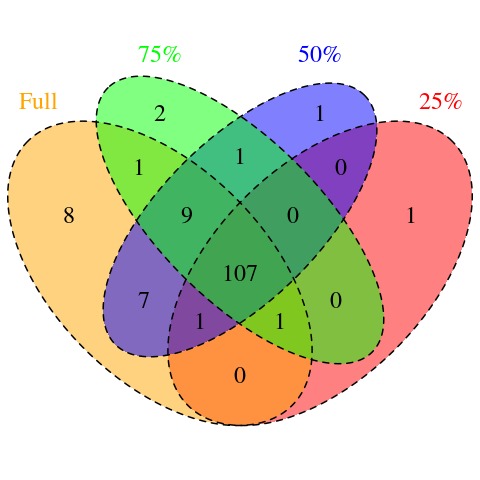}
\caption{BSmooth}\label{fig:BScov}
\end{subfigure}
\begin{subfigure}[b]{0.6\textwidth}
\includegraphics[width=0.6\textwidth]{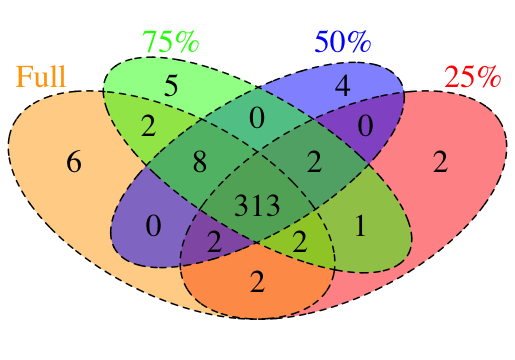}
\caption{MAGI}\label{fig:MAGIcov}
\end{subfigure}
\caption{Venn Diagram of Calls with Reduced Coverage. (a-b) The methods show broadly similar consistency to the M$^3$D method with respect to lower coverage.}
\end{figure}

Figures \ref{fig:BScov} and \ref{fig:MAGIcov} show the results of the test for reduced coverage for BSmooth and MAGI respectively.

\section{Expression Fold Changes}

\begin{figure}[!h]
\centerline{\includegraphics[width=0.7\textwidth]{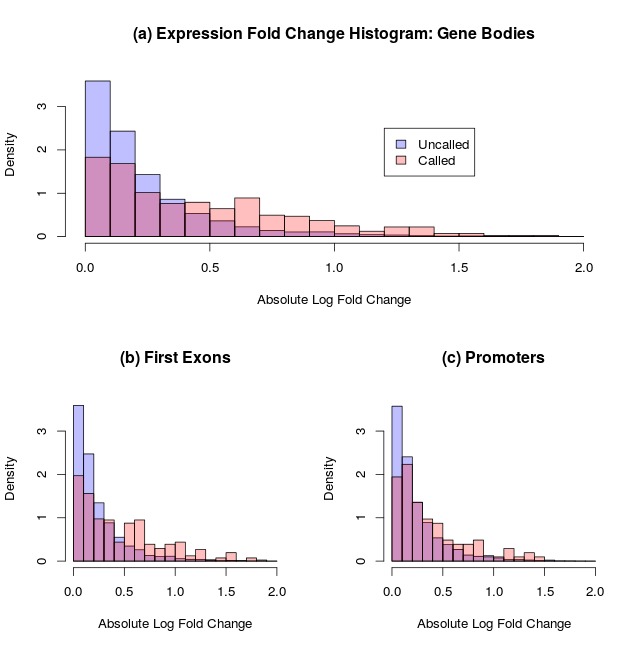}}
\caption{Absolute expression log-fold changes for called regions (red) and uncalled regions (blue) for (a) gene regions, (b) first exon regions and (c) promoter regions.}\label{fig:foldchange}
\end{figure}

Figure \ref{fig:foldchange} shows histograms for the absolute log-fold change of the called and uncalled regions for genes, first exons and promoters respectively. All three show an increase, but it is more pronounced for gene regions and first exons.

\begin{table}[ht]\caption{GO terms: Gene Regions}\label{table:geneRegions}
\centering
\begin{tabular}{rlll}
  \hline
 & ID & name & p.adjusted \\ 
  \hline
1 & GO:0071944 & cell periphery & 1.437708223467503E-25 \\ 
  2 & GO:0005886 & plasma membrane & 6.444067866565821E-22 \\ 
  3 & GO:0044459 & plasma membrane part & 4.062219312427493E-20 \\ 
  4 & GO:0032501 & multicellular organismal process & 3.1909227117764803E-13 \\ 
  5 & GO:0044425 & membrane part & 4.4083353990865556E-13 \\ 
  6 & GO:0003677 & DNA binding & 5.045748555333859E-13 \\ 
  7 & GO:0043565 & sequence-specific DNA binding & 9.122745220828804E-11 \\ 
  8 & GO:0044707 & single-multicellular organism process & 9.790514431161451E-11 \\ 
  9 & GO:0004872 & receptor activity & 1.6558910963111396E-10 \\ 
  10 & GO:0022610 & biological adhesion & 1.6465615260421583E-9 \\ 
   \hline
\end{tabular}
\end{table}

\begin{table}[ht]\caption{GO terms: First Exons}\label{table:FirstExonRegions}
\centering
\begin{tabular}{rlll}
  \hline
 & ID & name & p.adjusted \\ 
  \hline
1 & GO:0071944 & cell periphery & 1.2919365432790091E-5 \\ 
  2 & GO:0005886 & plasma membrane & 5.622471574689E-5 \\ 
  3 & GO:0003677 & DNA binding & 0.0012341786928683266 \\ 
  4 & GO:0032501 & multicellular organismal process & 0.004089560778137887 \\ 
  5 & GO:0044459 & plasma membrane part & 0.0067889743969307675 \\ 
  6 & GO:0045165 & cell fate commitment & 0.0067889743969307675 \\ 
  7 & GO:0044707 & single-multicellular organism process & 0.0067889743969307675 \\ 
  8 & GO:0007010 & cytoskeleton organization & 0.0067889743969307675 \\ 
  9 & GO:0022610 & biological adhesion & 0.00819729130007154 \\ 
  10 & GO:0051965 & positive regulation of synapse assembly & 0.015205197324440567 \\ 
   \hline
\end{tabular}
\end{table}

\begin{table}[ht]\caption{GO terms: Promoter Regions}\label{table:promRegions}
\centering
\begin{tabular}{rlll}
  \hline
 & ID & name & p.adjusted \\ 
  \hline
1 & GO:1990351 & transporter complex & 0.4262394723222016 \\ 
  2 & GO:0006351 & transcription, DNA-templated & 0.5120415178222222 \\ 
  3 & GO:0010977 & negative regulation of neuron projection development & 0.5261134605589649 \\ 
  4 & GO:0044459 & plasma membrane part & 0.6876369065744885 \\ 
  5 & GO:0003677 & DNA binding & 0.6876369065744885 \\ 
  6 & GO:0033764 & steroid dehydrogenase activity & 0.6876369065744885 \\ 
  7 & GO:0007260 & tyrosine phosphorylation of STAT protein & 0.6876369065744885 \\ 
  8 & GO:0006811 & ion transport & 0.6876369065744885 \\ 
  9 & GO:0019825 & oxygen binding & 0.6876369065744885 \\ 
  10 & GO:0080090 & regulation of primary metabolic process & 0.6876369065744885 \\ 
   \hline
\end{tabular}
\end{table}

\section{Gene Ontology Terms}

Tables \ref{table:geneRegions}, \ref{table:FirstExonRegions} and \ref{table:promRegions} contain the top ten enriched gene ontology terms for the H1-hESC cells vs K562 cell comparison study, for gene regions, first exons and promoter regions in Tables \ref{table:geneRegions}, \ref{table:FirstExonRegions} and \ref{table:promRegions} respectively. We include the Benjamini-Hochberg adjusted p-values. Note that none of the promoter region terms achieve statistical significance, while the gene and first exon terms show a high degree of overlap.


\end{document}